\documentclass[aps,pre,onecolumn]{revtex4}
\usepackage{epsfig}
\usepackage{amsmath}
\usepackage{graphicx,natbib}

\begin{document}

\title{Emergent complex neural dynamics: the brain at the edge\footnote{Nature Physics  6, 744-750 (2010)}}

\author{Dante R. Chialvo}

\affiliation{Department of Physiology,  David Geffen School of Medicine, UCLA,  Los Angeles, CA 90024, USA and Facultad de Ciencias Medicas, Universidad Nacional de Rosario, Rosario 2000, Argentina. }

\begin{abstract}

A large repertoire of spatiotemporal activity patterns in the brain is the basis for adaptive behaviour. Understanding the
mechanism by which the brainÕs hundred billion neurons and hundred trillion synapses manage to produce such a range of
cortical configurations in a flexible manner remains a fundamental problem in neuroscience. One plausible solution is the
involvement of universal mechanisms of emergent complex phenomena evident in dynamical systems poised near a critical
point of a second-order phase transition. We review recent theoretical and empirical results supporting the notion that the
brain is naturally poised near criticality, as well as its implications for better understanding of the brain.

\end{abstract}
 
\maketitle

Understanding the brain is among the most challenging problems to which a physicist can be attracted. As a system with an astronomical number of elements, each one known to have plenty of nonlinearities, the brain exhibits collective dynamics that in many aspects resemble some of the classic problems well studied in statistical physics. The contradiction, and the provoking point in these notes, is that only a minority of the publications in the field today are concerned with the understanding of the brain dynamics as a collective process. To the contrary, the great majority of the work explains the brain through explicit or implicit connectionist paradigms. In our opinion there is a need to reflect and recognize to what degree these collectivist-connectionist views imply more than just a semantic difference, and that its adequate resolution holds the key to resolve some of the more puzzling questions about the brain.
We review key results on emergent complex neural dynamics over the past few years. From the outset it should be noted that the intentionally provoking nature of these notes naturally induces a strong bias regarding cited publications; consequently this is neither a fair, nor historically correct, exhaustive or updated review of the relevant literature. Another cautionary note is that being a subject at the fringe of disciplines, physicists and biologists alike, will encounter boring passages on their most familiar topics. Nevertheless, for the sake of clarity, and with the forgiveness of the readers, we will proceed to (even excessively) define each issue at hand. 

\subsection*{What are the issues?}
Understanding human behavior and cognition requires the description of the laws of the underlying neural collective phenomena, the patterns of spatio-temporal brain activity. Formal approaches to study collective phenomena are one of the classical topics at the center of statistical physics, with recent novel and successful applications in diverse areas such as genetics, ecology, computer science, social and economic settings \cite{Bak97, Jensen,Buchanan,Christensen,Aldana,Balleza,Lux,Malamud,Nykter,Takayasu,Peters1,Peters2,Peters3}. While in all these fields there is a clear transfer of methods and ideas from statistical physics, a similar flow has only recently started to impact neuroscience. 

The main issue addressed here belongs to the Òunder the rugÓ class of problems in the field, namely, how the \emph{ very large} conglomerate of interconnected neurons produce a \emph{repertoire} of given behaviors in a \emph{flexible} and self organized way. Although colloquial explanations abound, when detailed models are constructed to account for that, each of the three \emph{emphasized} condiments are systematically violated. Either 1) the model is a low dimensional version of the neural structure of interest or 2) it produces a single behavior (hardwired in the system), and consequently 3) it cannot flexibly do anything else. It is only by arbitrarily changing the neuronal connections that most of current models can play a repertoire of behaviors. This requires a kind of supplementary brain holding the key to which connections need to be rewired. Of course, no proof of such supplementary brain has been offered, and this is the question that is screaming to be answered and seldom is even being asked.

Approaches to this problem, for a variety of historical and conceptual reasons, are still drawn from connectionist paradigms which restrict the dynamics to be generated by circuits, and consequently funnel our efforts in the same sterile direction. While collective properties have been mentioned for a long time, its relevance remains secondary to most of us. Even Hopfield's call \cite{Hopfield} three decades ago (in his ``Neural networks and physical systems with emergent collective computational abilities'' seminal paper), seems to have been forgotten, perhaps displaced by the appeal and initial excitement of computational ideas:

\emph{``Much of the architecture of regions of the brains of higher animals must be made from a proliferation of simple local circuits with well-defined functions. The bridge between simple circuits and the complex computational properties of higher nervous systems may be the spontaneous emergence of new computational capabilities from the collective behavior of large numbers of simple processing elements.''}

\subsection*{Emergence}
It is accepted that almost all macroscopic phenomena -from superconductivity to gravity and from economics to photosynthesis- are the consequence of an underlying collective dynamics of their microscopic components. In neuroscience, it is the macroscopic behavior (cognitive, emotional, motor, etc) aspect that will be ultimately understood as the emergent phenomena of an underlying neuronal collective. However, neurons being nonlinear elements, makes such understanding far from straightforward. It would be fair to say that while the problem is cast in terms most familiar to biology the solution is written in terms very familiar to physics.

Let's recall what emergent phenomena are. Emergence refers to the unexpected collective spatiotemporal patterns exhibited by \emph{large} complex systems. In this context, ``unexpected'' shows our inability (mathematical and otherwise) to derive such emergent patterns from the equations describing the dynamics of the individual parts of the system. As discussed at length elsewhere \cite{Bak97, Bak95}, complex systems are usually \emph{large} conglomerates of \emph{interacting} elements, each one exhibiting some sort of \emph{nonlinear} dynamics. Without entering into details, it is also known that the interaction can also be indirect, for instance through some mean field. Usually energy enters into the system, thus some sort of driving is present. The three \emph{emphasized} features, ( i.e., large number of interacting nonlinear elements)  are necessary, although not sufficient, conditions for a system to exhibit emergent complex behavior at some point.

As long as the dynamics of each individual element is nonlinear, other details are not important \cite{Bak97,anderson}; for instance, they can be humans, driven by food and other energy resources, from which some collective political or social structure eventually arises. Whatever the type of structure that emerges, it is unlikely to appear if one of the three above-emphasized properties is absent. For instance, it is well established that a small number of isolated linear elements are unable to produce unexpected behavior (indeed this is the case in which everything can be mathematically anticipated). 

\subsection*{Spontaneous brain activity is complex}
It is evident, from the very early electrical recordings a century ago, that the brain is spontaneously active, even in absence of external inputs. However obvious this observation could appear, it was only recently that the dynamical features of the spontaneous brain state started to be studied in any significant way. 

Recent work on brain rhythms at small and large brain scales showed that spontaneous healthy brain dynamics is not composed by completely random activity patterns or by periodic oscillations \cite{Buzsaki}. Careful analysis of the statistical properties of neural dynamics under no explicit inputs has identified complex patterns of activity previously neglected as background noise dynamics. The fact is that brain activity is always essentially arrhythmic regardless of how it is monitored, whether as electrical activity in the scalp (EEG), by techniques of functional magnetic resonance imaging (fMRI), in the synchronization of oscillatory activity \cite{Linkenkaer,Stam}, or in the statistical features of local field potentials peaks \cite{Plenz2007}.

It has been pointed out repeatedly \cite{Bullock,Logothetis, Eckhorn,Manning,Miller} that, under healthy conditions no brain temporal scale takes primacy over average, resulting in power spectral densities decaying of ``1/f noise''. Behavior, the ultimate interface between brain dynamics and the environment, also exhibits scale invariant features as shown in human cognition \cite{Gilden,Maylor,Ward} human motion \cite{Nakamura} as well as animal motion \cite{Anteneodo}. The origin of the brain scale free dynamics was not adequately investigated until recently, probably (and paradoxically) due to the ubiquity of scale invariance in nature \cite{Bak97}. Currently, there is increasing interest and the potential significance of a renewed interpretation of the brain spontaneous patterns is at least double. Its presence provides important clues about brain circuitÕs organization, in the sense that our previous ideas cannot easily accommodate these new findings. Also, the class of complex dynamics observed seems to provide the brain with previously unrecognized robust properties. These aspects will be reviewed, on two different scales, in the next sections.

\subsection*{Emergent complexity is always critical}
The commonality of scale-free dynamics in the brain naturally leads one to ask what physics knows about very general mechanisms able to produce such dynamics. Attempts to explain and generate nature's non- uniformity included several mathematical models and recipes, but few succeeded in creating complexity without embedding the equations with complexity. The important point is that  including the complexity in the model will only result in a simulation of the real system, without entailing any understanding of complexity. The most significant efforts were those aimed at discovering the conditions in which something complex emerges from the interaction of the constituting non-complex elements \cite{Bak97, Bak87}. Initial inspiration was drawn from work in the field of phase transitions and critical phenomena (see Box 1). Precisely, one of the novelties of critical phenomena is the fact that out of the short-range interaction of simple elements eventually long-range spatiotemporal correlated patterns emerge. As such, critical dynamics have been documented in species evolution \cite{Bak97}, ants collective foraging \cite{Beckers,Beekman} and swarm models \cite{Rauch}, bacterial populations \cite{Nicolis}, traffic flow in highways \cite{Bak97} and on the Internet  \cite{Takayasu}, macroeconomic dynamics \cite{Lux}, forest fires \cite{Malamud}, rainfall dynamics \cite{Peters1, Peters2, Peters3} and flock formation \cite{Cavagna}. Same rationale lead to the conjecture \cite{Bak97, Chialvo99,Bak2001} that also the complexity of brain dynamics was just another signature of an underlying critical process. Since at the point near the transition exist the largest number of metastable states, the brain can then be accessing the largest repertoire of behaviors in a flexible way. That view claimed that the most fundamental properties of the brain only are possible staying close to that critical instability (see Box 2) independently of how such state is reached or maintained.

\subsection*{Small scale: Cortical quakes} 
Beggs and Plenz \cite{Beggs2003,Beggs2004} reported the first convincing evidence that neuronal populations could exhibit critical dynamics. They described a novel type of electrical activity for the brain cortex called ``neuronal avalanches''. These collective neuronal patterns sit halfway in between two previously well-known cortical patterns: the oscillatory or wave-like highly coherent activity on one side and the asynchronic and incoherent spiking on the other.  Typically, each avalanche engages a variable number of neurons. What is peculiar is the statistical pattern that these avalanches follow. On the average, one observes many more small avalanches than big ones (e.g., each neuronal avalanche has a large chance to engage only a few neurons and a very low probability to spread and activate the whole cortical tissue (see Fig. 1)). In these experiments, a number of properties suggestive of criticality were estimated. This included a scale-free distribution of avalanche sizes following an inverse power law with an exponent close to 3/2 that agrees exactly with the theoretical expectation for a critical branching process, previously worked out by Zapperi et al\cite{Zapperi}. The avalanches lifetimeÕs statistics followed also an inverse power law with an exponent close to 2, which agrees with the theoretical expectation for a cascade of activity \cite{Eurich, Zapperi,Teramae}. 
\begin{figure}
\begin{center}
\includegraphics[width=.4\textwidth,angle=0,keepaspectratio,clip]{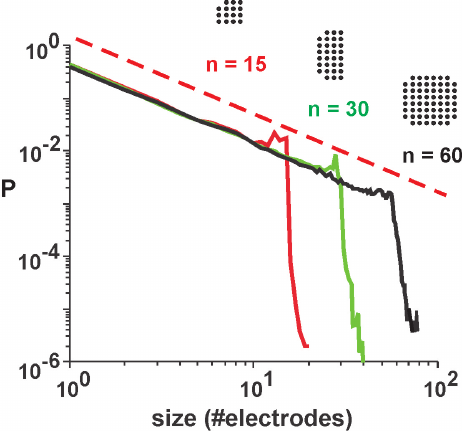}
\end{center}
\caption{Neuronal avalanches are complex. Size distribution of neuronal avalanches in mature cortical cultured networks follows a power law with an exponent close to 3/2 (dashed line) and exhibits finite size scaling. The relative probability of observing an avalanche covering a given number of electrodes is shown for three sets of grid sizes (insets with n=15, 30 or 60 sensing electrodes, equally spaced at $200\mu m$). The statistics is taken from data recordings lasting a total of 70 hours and accumulating 58000 ($+/-$ 55000) avalanches per hour (mean $+/-$ SD). (Re-plotted from Figure 4 of \cite{Chialvo2004}).}
 \label{fig1}
 \end{figure}

The initial scale invariance for the avalanches has already been replicated \cite{Mazzoni,Pasquale}. Furthermore, similar findings have been reported on a wide variety of diverse settings, including in vivo monkey's cortex \cite{Petermann} and adult cats \cite{Hahn}. In addition, the functional significance of the avalanches was highlighted by the fact that they were observed during the earliest time of the development of superficial layers in the cortex \cite{Stewart,Gireesh} requiring the presence of a neuro-modulator (dopamine) and a certain balance between excitatory and inhibitory transmission \cite{Beggs2003,Stewart,Gireesh}. 
 
The precise neuronal mechanisms leading to the observed scale-free avalanches is yet uncertain, despite modeling efforts underway (see Box 3), because similar statistics can be generated by several mechanisms other than critical dynamics. Yet, no convincing alternative experimental analysis or evidence has been presented up to now. Numerical evidence recently reported \cite{Touboul} suggesting non-critical alternatives gives inverse power law exponents one order of magnitude larger ($> 20$) than the 3/2 experimentally observed.
 
The stumbling block of the discussions concerning the origin of the avalanches has been the limitation to replicate only probability densities, either of sizes or durations. However, the debate can be placed in a more rigorous context if other invariants are analyzed. In that direction, recent results \cite{Plenz2009} provided novel experimental evidence for five fundamental properties of neuronal avalanches consistent with criticality. These were: (1) time scales separation between the dynamics of the triggering event and the avalanche itself. This is demonstrated by the fact that the inverse power law density of avalanches sizes and lifetimes remain invariant to slow driving; (2) stationary of avalanches size statistics despite wide avalanching rate fluctuations, excluding non-homogeneous Poisson processes; (3) the avalanche probabilities preceding and following main avalanches obey Omori's law for earthquakes; (4) the average size of avalanches following a main avalanche decays as an inverse power law; (5) avalanches spread spatially on a fractal. Overall, these results support the notion that neuronal avalanches are the manifestation of criticality in the brain and exclude, in some cases explicitly, the majority of the mechanisms discussed in the literature as alternatives to criticality.

\begin{figure}
\begin{center}
\includegraphics[width=1\textwidth,angle=0,keepaspectratio,clip]{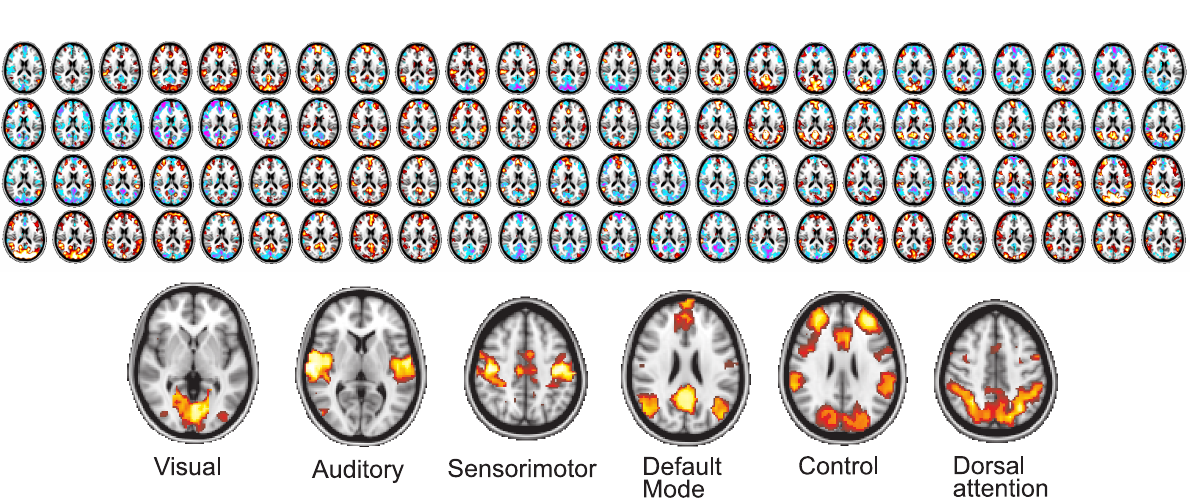}
\end{center}
\caption{Large scale emergent brain networks: The analysis of interactions during spontaneous human brain spatiotemporal patterns reveals emergent networks. Top sequence of images shows in red/blue the increases/decreases over the mean fMRI BOLD during four minutes of consecutive brain resting (Single-subject consecutive data, starting from the top-left corner where each row is 1 minute, and images are taken at 2.5 sec. intervals). The bottom images are the results of computing linear correlations between the activity of a small region within the networks of interest and the rest of the brain (stronger correlations indicated with brightest colors). These networks correspond to the six mayor systems in the brain: visual, auditory, sensorimotor, default mode, executive control and dorsal attention (From Chialvo, unpublished data. The MNI152 coordinates for the slices are z=18  for the top sequence and z=0,8,44,24,26,44 for the bottom figures, left to right).}
 \label{fig2}
 \end{figure}

\subsection*{Large scale} 
Probably the first report concerning mesoscopic patterns in connection with behavior was Kelso et al \cite{Kelso92} brain imaging analysis pursuing their previous observation that human hand movements exhibit abrupt phase transitions for increasing cycling frequency \cite{Kelso84}. They were able to show, using magneto-encephalographic techniques, spontaneous transitions in neuromagnetic field patterns in the human brain. These transitions happened at a critical value of a systematically varied behavioral parameter supporting ``the thesis that the brain is a pattern forming system that can switch flexibly from one coherent state to another''.   Similar considerations and concerns were expressed in these early days, by commenting that: ``In summary, higher brain functions in humans such as perception, learning and goal directed movement are often hypothesized to depend on the collective dynamics of large numbers of interacting neurons distributed throughout the cortex. But typical signs of cooperative phenomena are not accessible through single neuron investigations. On the other hand, from studies of nonequilibrium systems it is well known that at critical points, spatial and temporal patterns form in a so-called self-organized fashion'' \cite{Kelso92}.
 
A large body of work needs to be omitted here to be able to fast forward to present day, when it is recognized that the brain is spontaneously creating and reshaping complex functional networks of correlated dynamics responding to the neural traffic between regions. These networks had been recently studied, using functional magnetic resonance imaging in humans. The flurry of activity in this area could be well gauged by the words chosen by the author \cite{Werner} of a recent review stating: ``Commenting on the wealth of existing data on anatomical and functional cortical networks organization may seem like ``carrying coals to Newcastle''. Extensive reviews \cite{Bullmore,Sporns} cover the statistical physics approaches that are increasingly being use to analyze this large body of complex data. 
 
For the purpose of this note, it is relevant to limit our attention to the study of spontaneous ``resting'' fMRI dynamics. \cite{Fox,Smith}. Brain ``rest'' is defined -more or less unsuccessfully- as the state in which there is no explicit brain input or output, or overt external stimulation. The subject is scanned while lying with eyes closed, and instructed to avoid falling  asleep.  Each of the thousands of signals (so called BOLD, for blood oxygenation level dependent) obtained from these experiments reflects the amount of neural activity on each small region of typically dozen of cubic millimeters, allowing one to map the entire activity of the brain. An example is presented as a sequence in Figure 2, which shows for graphical purposes only one of the many slices that are recorded. From careful visual inspection of the data it appears already that there are important spatiotemporal correlations, somehow resembling the image of passing clouds. The fascinating point here is that from rather simple linear cross correlations of the BOLD signals a few  collective groups emerge. This is shown by the clusters in the bottom panels, which were found to closely match the same regions responding to a wide variety of different activation conditions \cite{Fox,Smith}. Thus, at rest the ``passing clouds'' (i.e., the collective spatiotemporal dynamics) visit the same brain regions that are activated during any given active behavior. The relevance of these findings is further highlighted by the fact that these networks are identifiable with great consistency across subjects \cite{Xiong,  Cordes, Beckmann}, even during sleep \cite{Fukunaga} or anesthesia \cite{Vincent}. 
 
A natural question arising at this point is what kind of known dynamical scenario corresponds to these brain resting patterns. This was tackled initially in three recent reports. In the first Kitzbichler et al \cite{Kitzbichler} analyzed functional MRI and MEG data recorded from normal volunteers at resting state using phase synchronization between diverse spatial locations. They reported a scale invariant distribution for the length of time that two brain locations on the average remained locked. This distribution was also found in the Ising and the Kuramoto model \cite{Kuramoto} at the critical state.  
 
In the second report Fraiman et al \cite{Fraiman} compared the paradigmatic 2-dimensional Ising model at various temperatures with the resting brain data. Correlations networks were prepared by computing cross-correlation between the Ising states at all lattice points and placing links between those points (nodes) with correlation larger than a certain threshold. Similar computations were conducted for the brain fMRI data. After comparing the most descriptive networks properties, the authors concluded that while the Ising networks at sub- and supercritical temperatures greatly differ from the brain networks, those derived at the critical temperature are ``indistinguishable from each other''. The example shown in Figure 3, shows one of the properties compared, the distribution of the number of links (i.e., degree) for these networks. Notice the fat tails in the brain data, (reported earlier in refs\cite{Eguiluz,van}) which are only replicated when the Ising model is posed at critical temperature.  In addition, calculation of the fraction of sites with positive and negative correlations, (a variable related to magnetization elsewhere) showed values close to one, in agreement with previous results by Baliki et al \cite{Baliki} which suggested this balance as an index of healthy  brain function at rest. Overall, these results show that networks derived from correlations of fMRI signals in human brains are indistinguishable from networks extracted from Ising models at critical temperature.
 
The third effort directed to shed light on the mechanism underlying resting fMRI dynamics is due to Expert et al \cite{Expert} who examined the two point correlation function after successive steps in spatial coarse graining, a renormalization technique widely used in critical phenomena. Their results show spatial self-similarity which in addition to temporal 1/f frequency behavior of the power spectrum are indicative of critical dynamics. 

Since the initial fMRI work \cite{Eguiluz, van,Salvador} progress has been made to use these approaches to evaluate the integrity of brain function under normal \cite{Damoiseaux}  and pathological conditions \cite{Broyd,Zhang} including Alzheimer \cite{He}, squizophrenia \cite{Garrity}  and epilepsy \cite{Laufs,milton}.   Even the impact of long enduring chronic pain seems to alter brain dynamics beyond the feeling of pain itself \cite{Baliki}, thus motivating further work to better understand the fundamental mechanisms behind the brain resting state  large-scale organization.
\begin{figure}
\begin{center}
\includegraphics[width=.5\textwidth,angle=0,keepaspectratio,clip]{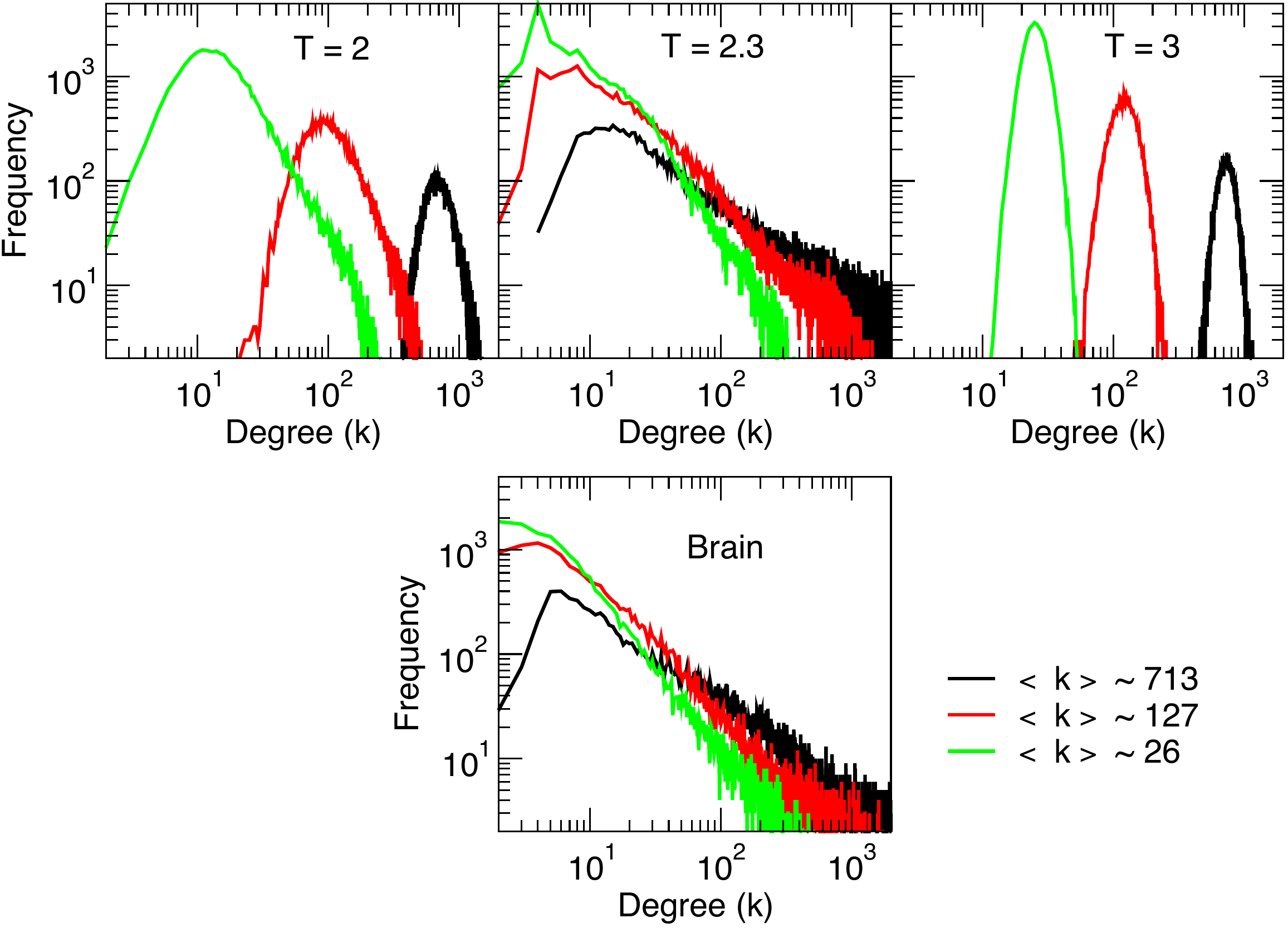}
\end{center}
\caption{Complex networks derived from the brain fMRI data mimic those from the Ising model only at the critical temperature. Plots correspond to the degree (k) distribution of the network derived from the fMRI brain resting data (bottom panel) and from the Ising model (top three panels) at T = 2, T = 2.3 and T = 3, for different values of averages degrees (From Fraiman et al \cite{Fraiman}).
}
 \label{fig3}
 \end{figure}

\subsection*{Summary and Outlook}
We have reviewed recent key results of emergent collective neural dynamics.  It is important to note that, at present time, no theory (in the sense of the initial paragraphs) comprehensibly can accommodate these results without invoking criticality. One motivation for neuroscience to look at the physical laws governing other complex systems is the hope that universality will give the field an edge. Instead to search for ad-hoc laws for the brain, under the pretense that biology is special, most probably a good understanding of universal laws might provide a breakthrough since brains must share some of the fundamentally laws of nature. A major difference between the preceding decade and now is that, as presented in this note, there is spatiotemporal brain data to confront theories with; a playground awaiting for physicists to take up the challenge of explaining the underlying mechanism of the collective.

\subsection*{Box 1: What is special about being critical?}
 Given the claims that brain dynamics can be critical, let's recall what is special about such a state. The generality of the scenario of ferromagneticÐparamagnetic phase transition is used, without implying that the brain would reach criticality in this way. As an iron magnet is heated, the magnetization decreases until it reaches zero beyond a critical temperature Tc. Individual spins orientations are, at high temperatures, changing continuously in small groups. As a consequence, the mean magnetization, expressing the collective behavior, vanishes. At low temperature the system will be very ordered exhibiting large domains of equally oriented spins, a state with negligible variability over time. In between these two homogeneous states, at the critical temperature Tc, the system exhibits very different properties both in time and space. The temporal fluctuations of the magnetization are known to be scale invariant. Similarly, the spatial distribution of correlated spins show long-range (power law) correlations. It is only close enough to Tc that large correlated structures (up to the size of the system) emerge, even though interactions are with nearest neighbor elements. In addition, the largest fluctuations in the magnetization are observed at Tc. At this point the system is at the highest susceptibility, a single spin perturbation has a small but finite chance to start an avalanche that reshapes the entire systemÕs state, something unthinkable on a non-critical state. Many of these dynamical properties, once properly translated to neural terms, exhibit striking analogies to brain dynamics. Neuro-modulators, which are known to alter brain states acting globally over nonspecific targets, could be thought as control parameters, as is temperature in this case.

 \begin{figure}
\begin{center}
\includegraphics[width=.5\textwidth,angle=0,keepaspectratio,clip]{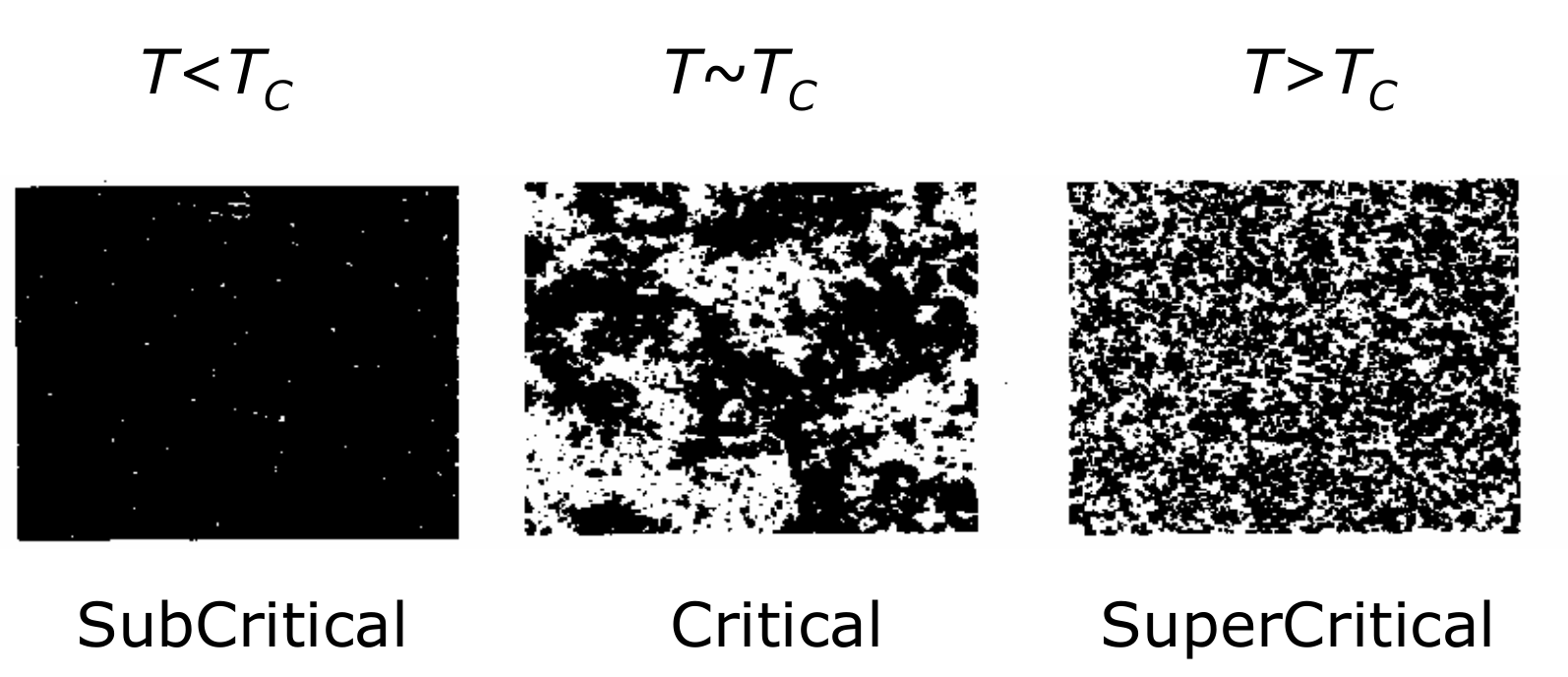}
\end{center}
\caption{Complex is critical: Three snapshots of the spin configurations at one moment in time for three temperatures (subcritical, critical and supercritical) from numerical simulations of the Ising model (d=2). Only at the critical temperature systems exhibiting a second-order phase transition show the highly heterogeneous correlated domains commonly seen as complex, while both sub- and supercritical conditions result in homogeneous states.
}
 \label{figbox}
 \end{figure} 
 
\subsection*{Box 2: Why we need a brain at all?}
It is self-evident that the brains we see today are those that inherited an edge useful to survive. In light of this, how consistent with Darwinian constraints could it be to suggest that the brain should evolve to be near a critical point? The answer, in short, is that brains should be critical because the world in which they must survive is to some degree critical as well. Lets see the alternatives: in a subcritical world, everything would be simple and uniform (as in the left panel of Box 1) and there would be nothing to learn; a brain would be completely superfluous. At the other extreme, in a supercritical world, everything would be changing continuously (as in the right panel of Box 1); under these circumstances there would not be sufficient regularity as to make learning possible or valuable. Thus, brains are only needed to navigate a complex, critical world, where surprising events still have a finite chance of occurring.
 In other words, animals need a brain because the world is critical \cite{Bak97, Bak2001,Bak87, Bak95}. Furthermore, a brain not only has to remember, but also has to forget and adapt. In a subcritical brain, memories would be frozen. In a supercritical brain, patterns change continuously so that no long-term memory would be possible. To be highly susceptible, the brain itself has to be in an in-between, critical state. 
 
Which generic features of systems at criticality should be expected in brain experiments?
 
1. At relatively large scale:
 
Cortical long-range correlations in space and time, 
 
Correlation length divergence
 
Near-zero magnetization or equivalently, the presence of anti-correlated cortical states.
 
2. At relatively small scale:
 
Cortical circuits exhibit ÔÔneuronal avalanchesÕÕ, cascades of activity obeying inverse power law statistics as well as long-range correlations. 
 
3. At behavioral level:
 
Adaptive human behavior should be ÔÔburstyÕÕ appearing unstable, as it was always at the ``edge of failure''. 
 
Life-long learning continuously ``raises the bar'' to more challenging tasks, making performance critical as well.   

\subsection*{Box 3: Models}
At small scale, explicit models have been already presented in which criticality can be self organized either by some form of Hebbian learning \cite{Bienenstock,magnasco} or by the inclusion of activity-dependent depressive synapses \cite{Levina}. Some novel properties endowed by criticality have been also studied, such as the widest dynamical range and optimal sensitivity to sensory stimuli shown by Kinouchi and Copelli \cite{Kinouchi}. Concerning learning, de Arcangelis and Herrmann \cite{deArcangelis} extending Bak \& Chialvo \cite{Bak2001,Chialvo99} earlier network model, found recently that avalanches of activity are able to shape a network with complex connectivity as well as learning logical rules. A handicap of these models is the absence of ongoing activity, an important brain feature. This aspect is covered by Marro and colleagues \cite{Marro} networks which  show spontaneous  unstable dynamics and nonequilibrium phases in which the global activity wanders irregularly among attractors resulting in 1/f noise as the system falls into the most irregular behavior. 
 At large scale although brain data sets with unprecedented spatiotemporal resolution are now available, there is no model able to mimic a phase transition at such brain scale. Nevertheless, the challenge to construct data driven models at this level is being taken, as shown by Honey et al \cite{Honey} recent efforts to model fMRI resting state.

 \section*{Acknowledgments}
DRC is with the National Research and Technology Council (CONICET) of Argentina. This work was supported by CONICET and by NIH NINDS NS58661. Thanks to E. Tagliazucchi and F. Lebensohn-Chialvo for reading the manuscript.

\end{document}